\documentclass[prx,showpacs,superscriptaddress,floatfix,twocolumn]{revtex4-1}

\usepackage{array,multirow,graphicx}
\usepackage{braket}
\usepackage{epstopdf}
\usepackage{graphicx}
\usepackage{siunitx}
\usepackage{nicefrac}
\usepackage[normalem]{ulem}
\usepackage{varioref}
\usepackage{xcolor}
\usepackage{xfrac}
\usepackage{xr-hyper}
\usepackage{hyperref}
\usepackage{amsmath}
\usepackage{amssymb}
\usepackage{amsfonts}
\usepackage{mathptmx}

\begin{document}

\title{Pattern Formation by Electric-field Quench in Mott Crystal}

\author{Nicolas~Gauquelin}
\affiliation{Electron Microscopy for Materials Research (EMAT), Department of Physics, University of Antwerp,  BE-2020 Antwerpen, Belgium \\
NANOlab Center of Excellence, University of Antwerp, BE-2020 Antwerpen, Belgium}
\author{Filomena~Forte}
\affiliation{CNR-SPIN, I-84084 Fisciano, Salerno, Italy \\
Dipartimento di Fisica "E.R.~Caianiello", Universit\`{a} di Salerno, I-84084 Fisciano, Salerno, Italy}
    
\author{Daen~Jannis}
\affiliation{Electron Microscopy for Materials Research (EMAT), Department of Physics, University of Antwerp,  BE-2020 Antwerpen, Belgium \\ NANOlab Center of Excellence, University of Antwerp, BE-2020 Antwerpen, Belgium}
   
\author{Rosalba~Fittipaldi}
\affiliation{CNR-SPIN, I-84084 Fisciano, Salerno, Italy \\ 
Dipartimento di Fisica "E.R.~Caianiello", Universit\`{a} di Salerno, I-84084 Fisciano, Salerno, Italy}

\author{Carmine~Autieri}
\affiliation{International Research Centre MagTop, Institute of Physics, Polish
Academy of Sciences,~\\
Aleja Lotnik\'ow 32/46, PL-02668 Warsaw, Poland}

\author{Giuseppe~Cuono}
\affiliation{International Research Centre MagTop, Institute of Physics, Polish
Academy of Sciences,~\\
Aleja Lotnik\'ow 32/46, PL-02668 Warsaw, Poland}

\author{Veronica~Granata}
\affiliation{Dipartimento di Fisica "E.R.~Caianiello", Universit\`{a} di Salerno, I-84084 Fisciano, Salerno, Italy}
    
\author{Mariateresa~Lettieri}
\affiliation{CNR-SPIN, I-84084 Fisciano, Salerno, Italy}

\author{Canio~Noce}
\affiliation{Dipartimento di Fisica "E.R.~Caianiello", Universit\`{a} di Salerno, I-84084 Fisciano, Salerno, Italy \\
CNR-SPIN, I-84084 Fisciano, Salerno, Italy}

\author{Fabio~Miletto Granozio}
\affiliation{CNR-SPIN, I-80126 Napoli, Italy \\ 
Dipartimento di Fisica, Universit\`{a} di Napoli, Napoli, Italy}

\author{Antonio~Vecchione}
\affiliation{CNR-SPIN, I-84084 Fisciano, Salerno, Italy \\
Dipartimento di Fisica "E.R.~Caianiello", Universit\`{a} di Salerno, I-84084 Fisciano, Salerno, Italy}

\author{Johan~Verbeeck}
\affiliation{Electron Microscopy for Materials Research (EMAT), Department of Physics, University of Antwerp,  BE-2020 Antwerpen, Belgium \\
NANOlab Center of Excellence, University of Antwerp, BE-2020 Antwerpen, Belgium}

\author{Mario~Cuoco}
\affiliation{CNR-SPIN, I-84084 Fisciano, Salerno, Italy \\
Dipartimento di Fisica "E.R.~Caianiello", Universit\`{a} di Salerno, I-84084 Fisciano, Salerno, Italy}



\begin{abstract} 
The control of Mott phase is intertwined with the spatial reorganization of the electronic states.
Out-of-equilibrium driving forces typically lead to electronic patterns that are absent at equilibrium, whose nature is however often elusive. 
Here, we unveil a nanoscale pattern formation in the Ca$_2$RuO$_4$ Mott insulator.
We demonstrate how an applied electric field spatially reconstructs the insulating phase that, uniquely after switching off the electric field, exhibits nanoscale stripe domains. 
The stripe pattern has regions with inequivalent octahedral distortions that we directly observe through high-resolution scanning transmission electron microscopy. The nanotexture depends on the orientation of the electric field, it is non-volatile and rewritable.
We theoretically simulate the charge and orbital reconstruction induced by a quench dynamics of the applied electric field providing clear-cut mechanisms for the stripe phase formation. 
Our results open the path for the design of non-volatile electronics based on voltage-controlled nanometric phases.
\end{abstract}

\maketitle


\noindent There are various paths to drive a changeover of the Mott insulating state~\cite{Mott1961} by either applying pressure or strain, changing the temperature nearby the Mott transition or doping the system away from integer filling, corresponding to bandwidth, temperature and density control, respectively ~\cite{Georges1996,ImadaRMP1998,Werner2007,Ahn2003,Takagi2010,Zitko2007,Ricco2018}. 
{{The resulting phenomena have broad impact in condensed matter physics for both fundamental~\cite{Georges1996,ImadaRMP1998} and technological perspectives~\cite{Ahn2003,Takagi2010}}}.

There are two scenarios that are often encountered in proximity of Mott phases: i) the
occurrence of superconductivity when the insulating phase is destroyed, as for the emblematic case of cuprates~\cite{Lee2006} with
magnetism playing an important role too, and ii) the tendency to form inhomogeneous electronic patterns due to the first order character of the Mott transition and the competing length scales of localized and itinerant electronic degrees of freedom~\cite{Zaanen1989,Low1994,dagotto2003}.
Recently, it has been pointed out that the application of an electric field, both static or dynamic, can be an ideal knob to control the conducting properties of correlated materials by inducing insulator-to-metal transitions and novel quantum phases of matter~\cite{Gianetti2016,Cavalleri2018,Oka2019,DelaTorre2021}.  
Depending on its amplitude, an applied gate voltage can yield a dielectric breakdown and electronic avalanches~\cite{Vaju2008} or activate collective low-energy lattice and spin-orbital excitations~\cite{Okazaki2013}.
The transitions which emerge from the Mott insulating state can hence involve multiple degrees of freedom and be marked or not by significant changes in the crystal structure, as is the case for V$_2$O$_3$~\cite{Ronchi2022},  VO$_2$~\cite{Kim2010,Wen2013,Zimmers2013,Hsieh2014} and Fe$_3$O$_4$ systems~\cite{Burch2014}.  
In this framework, CRO represents a paradigmatic material platform to assess the interplay of electron correlations and electron-lattice coupling in the presence of multi-orbital physics~\cite{AlexanderPRB1999,NakatsujiPRL00,DasPRX2018} with spin-orbit and Hund’s interactions ~\cite{Sutter2017}. Indeed, CRO is a Mott-insulator at room temperature and on heating through T$_{\text{MI}}$=83~°C it undergoes an insulator-to-metal transition accompanied by an abrupt structural change~\cite{Alexander1999}, without varying the crystal symmetry. The structural transition involves an orbital reconstruction from a preferential out-of-plane orbital occupancy of the 4d states ($xz,yz$) to a dominant orbital configuration with in-plane character ($xy$). At lower temperature, this redistribution turns into an orbitally ordered state~\cite{Okazaki2013,Porter2018}. The application of electric field, through current and optical pulses, or {{the use of thermal quench}} have been shown to melt the insulating phase~\cite{Nakamura2013,Cirillo2019,Zhang2019,Vitalone2022} resulting into the formation of phase coexistence ~\cite{Zhang2019},including nanometric regions~\cite{Vitalone2022} at the boundaries of micrometric domains having metallic and insulating character. Nevertheless, while spatial inhomogeneities can form and inequivalent structural components compete, the origin and the mechanisms for the spatial reorganization remain mostly unexplained.

The problem of domains formation is particularly challenging in the context of Mott transitions as they are first order type in real materials. Domains formation is a general and complex phenomenon~\cite{Seul1995, Portmann2003, Yu2012}, with modulations that often result from the competition between short-range attractive and long-range repulsive interactions.
In dynamical conditions, domains can arise from the quench of the interactions or by quenching the temperature from above to below the ordering transition \cite{Bray1994}. 
Whether similar reorganization phenomena after electric or orbital quench can be encountered in correlated systems exhibiting insulator-to-metal transition is an outstanding problem not yet fully uncovered. 

In this manuscript, we face this challenge and unveil a novel path to induce {{as well as turn on and off}} the formation of nanotextures by means of an applied electric field in a Mott insulator, focusing on the case of the CRO system. The emergent phase remarkably depends on the electric field orientation. It is a stable configuration that can be erased by voltage or temperature and regenerated with the same voltage quench protocol. The formation of the domains is ascribed to a nontrivial orbital dynamics that is activated by the electric field. 
We demonstrate that the electric field is able to affect and reduce the orbital population unbalance among the $xy$ and $(xz,yz)$ states. 
The nontrivial {{orbital}} relaxation allows for the formation of interfaces of long and short octahedra. 
We show that these interfaces are electrically active, thus they can interact by stabilizing a stripe pattern.   



\begin{figure*}[h!]
\includegraphics[width=0.75\textwidth]{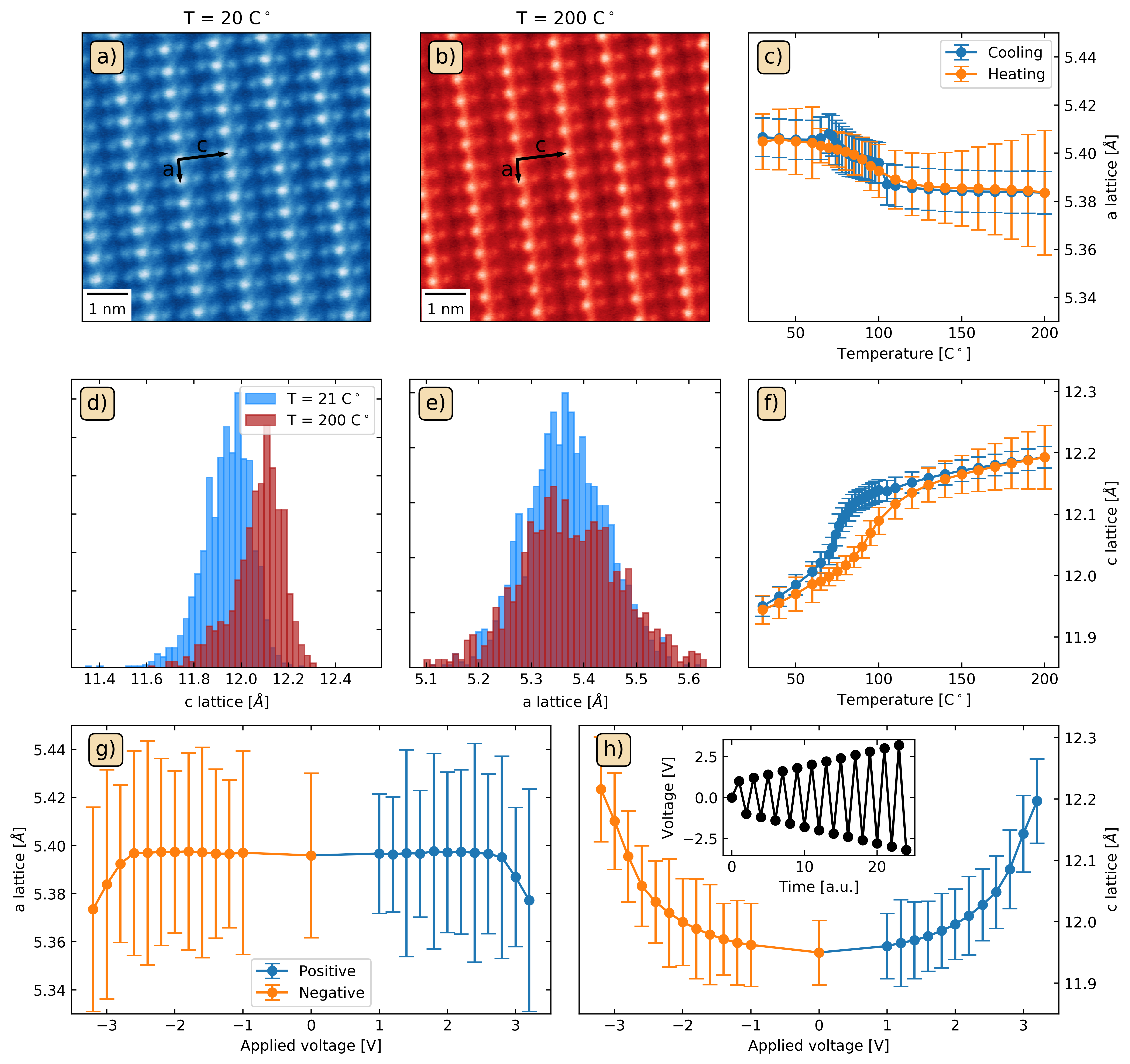}
\caption{High Angle Annular Dark Field (HAADF-STEM) image from the \textbf{(a)} low and \textbf{(b)} high temperature structural phase of CRO, respectively. \textbf{(d)} The histogram of the $c$ lattice constant for low and high temperature. The lattice constant is determined by fitting the atomic positions of the Ru atoms with a 2d Gaussian function and from this the lattice constant can be calculated. \textbf{(e)} Similar to \textbf{(d)} but the $a$ lattice constant is shown. \textbf{(c,f)} The evolution of the two lattice parameters as a function of temperature when heating and cooling the specimen. The lattice constants are determined from the NBED experiments. \textbf{(g,h)} The lattice constants as a function of the applied voltage. In the inset of panel \textbf{(h)} the sequence of applied voltage is shown. The setup of the contacts between the electrodes and the sample is reported in the Extended Figure 1. The applied voltage leads to an electric field that is oriented along the $b$ axis.}
\label{fig:fig1}
\end{figure*}

Let us start by considering the structural evolution across the thermally induced insulator-to-metal transition.  Nakamura et al.~\cite{Nakamura2013} reported the change of the lattice parameters as measured upon heating of a bulk specimen. Hence, to set the reference, the first issue we aim to address is how the impact of the thermal effects on the structure manifests at the nanoscale.  
Figure \ref{fig:fig1}a and Figure \ref{fig:fig1}b compare the High-Resolution Transmission Electron Microscopy High Angle Annular dark field (HRSTEM-HAADF) image of the CRO sample taken at room temperature (20°C) and at a representative higher temperature (200°C). Note that we do not have access to the b-axis as it is in the direction of the electron beam. Although the change of lattice parameter from the short (S-) to the elongated (L)-phase is almost invisible by figure inspection, the fitting of the Ru atomic columns with a 2D gaussian distribution can provide the amplitude of the in-plane and out-of-plane lattice parameters. We find an expansion of the $c$ lattice parameter from 11.9 to 12.2 {\AA}, as shown in Figure \ref{fig:fig1}d, while the variation of the in-plane ($a$ lattice parameter) is, on the other hand, small (Figure \ref{fig:fig1}c) in amplitude, with almost no change in the histogram displayed on Figure \ref{fig:fig1}e. This analysis indicates that, by increasing the temperature, the RuO$_6$ octahedra and the unit cell elongate.
Nanobeam electron diffraction (NBED) was used to locally determine the lattice parameter. This method collects a diffraction pattern, from the transmitted electrons, at each probe position while raster scanning the nanometer electron beam over the specimen. The information on the lattice parameters comes from the position of the peaks in the diffraction pattern, which is collected at each probe position. The peaks in a diffraction pattern arise from the constructive interference of the electron wave coming from the interaction of the electron with crystal. The position of the diffraction peaks can be translated into a scattering angle from which the lattice parameters can be determined via the use of Bragg's law. Since the transmitted electrons are collected, the measured lattice parameters arise from the entire thickness of the specimen ($\sim$~100~nm). These diffraction patterns were collected along the [010] zone axis in the case of Figure \ref{fig:fig1}.
Figure \ref{fig:fig1}c and Figure \ref{fig:fig1}f report the evolution of the $a$ and $c$ lattice parameters as measured in-situ from the diffraction pattern every 10~°C. The analysis clearly signals the phase transition and its hysteresis loop of almost 30°C between heating and cooling cycle, which is consistent with the previously reported findings~\cite{Nakamura2013}. 

Interestingly, the application of a constant electric current induces the presence of a domain structure with stripes at the interface between a metallic and an insulating domain ~\cite{Zhang2019}.
Although inhomogeneities occur, in Ref.~\cite{Mattoni2020} it is shown that locally a metal-to-insulator changeover always occurs at the same transition temperature, irrespective of being driven by temperature or by current. These phenomena thus indicate a tied connection between thermal and electric current driven effects. 
Here, we aim to verify whether the electronic phase reconstruction happens or not at the nanoscale, in the presence of an applied electric field when the electric current is not allowed to flow through the specimen. The experiment is performed by employing a capacitor like geometry, as described in the supplementary material, for a specimen where the electric field is applied along the a-crystallographic axis. In this experimental configuration, the voltage has been increased with a saw-tooth profile having the following sequence: 0;+1 V;-1 V;+1.2 V;-1.2 V...+3.2 V;-3.2 V (see inset in Figure \ref{fig:fig1}h). Such voltage variation leads to a progressive increase of the $c$ lattice constant as a function of the applied voltage accompanied by an almost unchanged behavior of the $a$ lattice parameter amplitude. Here, two relevant remarks are in place: 1) the lattice parameter measured at a specific voltage does not depend on the orientation (sign) of the electric field but solely on its amplitude, 2) the lattice parameter value achieved at 3.2 V (corresponding to approximately 3.2 kV/cm), at room temperature, has the same value of that found at 200°C at zero voltage. This is also consistent with our measured bulk value and with published data ~\cite{Nakamura2013}. 
Additionally, the analysis of the spatially resolved maps demonstrates that the distribution of the lattice parameter is uniform and, thus, no pattern formation in real space is observed at any applied voltage.


An unexpected and different behaviour is instead achieved when the voltage is quenched to zero from the maximum voltage configuration. This observation is obtained in the devised capacitor geometry (see supplementary material), where the Joule heating is expected to play a minor role compared to the electric current flow setup~\cite{Cirillo2019,Mattoni2020,Zhang2019}.

Single crystalline specimens of the bulk single crystal of CRO were prepared using the focused ion beam (FIB) and attached to one of the capacitor plates to have the electric field applied along the three different crystallographic directions, as shown in the supplementary material.
For this setup, distinct spatial patterns are observed depending on the direction of the applied electric field before the quench to zero voltage. 

\begin{figure*}
\includegraphics[width=0.8\textwidth]{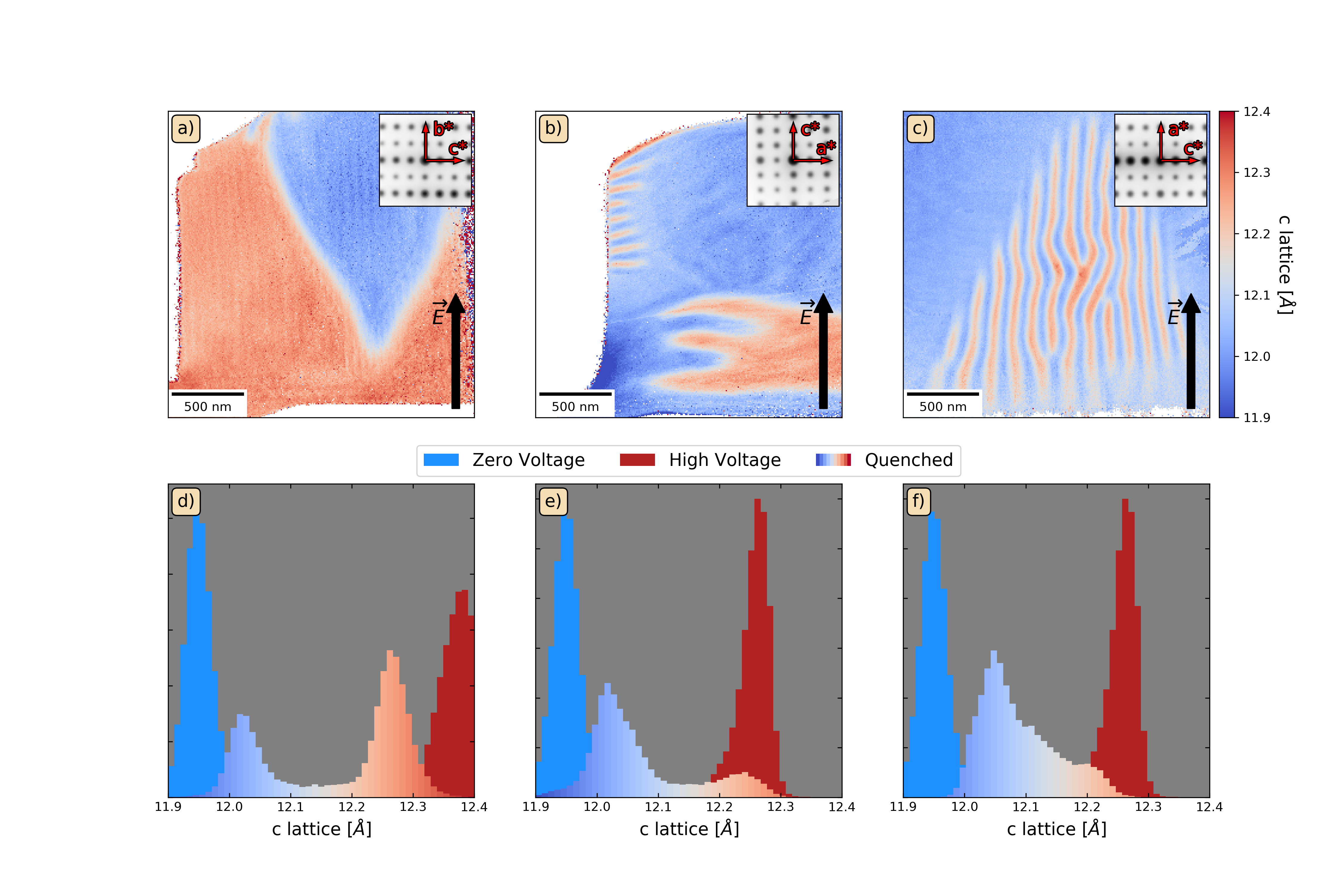}
\caption{\textbf{(a-c)} The real space map of the $c$ lattice parameter after the voltage is quenched to zero amplitude for three different orientation of the electric field. The orientation of the crystal with respect to the electric field is indicated in each panel. In the inset images, the average diffraction pattern is shown. 
\textbf{(d-f)}  The histogram of the lattice parameter at zero voltage and maximum applied voltage indicates the distribution of the lattice parameters amplitude for the corresponding voltage configurations. We find that after the quench the distribution exhibits a bimodal lineshape that reflects the occurrence of stripes or domains with unit cells having short and long $c$ lattice parameters. The geometry of the electrical contacts corresponds to an open circuit with the sample gated only on one side. The details of the electrical setup is reported in the Extended Figure 1. The bottom side in the panels a-c corresponds to the region of the contact of the sample with the electrode through which the electric field is applied. The white region on the left side of the panels (a) and (b) refers to the interface with the vacuum at the boundary of the sample.}
\label{fig:fig2}
\end{figure*}

For clarity only the $c$ lattice parameter is reported as it gives the most significant variation. When the electric field is applied along the $b$ crystallographic direction, two large domains are observed (Figure \ref{fig:fig2}a) with short c-axis in the center of the specimen and longer c-axis in the regions which are closer to the gate contact and the surface, i.e. along the bottom and left side of the Figure \ref{fig:fig2}a, respectively. 
Figure \ref{fig:fig2}d (similarly to panels e and f) represents the histogram of the $c$ lattice parameters corresponding to 0 voltage, maximum voltage and quenched state as shown in the supplementary material. 
When the applied voltage induces an electric field along the $c$-crystallographic direction and then is quenched to zero, stripes with a periodic sequence of regions with short and long octahedra appear along the c-direction of the specimen, as shown in Figs. \ref{fig:fig2}b and \ref{fig:fig2}e. The direction of the stripe modulation is perpendicular to the electric field orientation. On the other hand, when the voltage is applied along the $a$-crystallographic axis (Figures \ref{fig:fig2}c and \ref{fig:fig2}f), stripes are observed parallel to the applied electric field and mostly close to the interface between the sample and the electrode corresponding to the $ac$ plane.
Note that this stripe configuration has both a different generating mechanism and a diverse orientation when compared to the findings in Ref. \cite{Zhang2019}.

\begin{figure*}[t]
\includegraphics[width=0.8\textwidth]{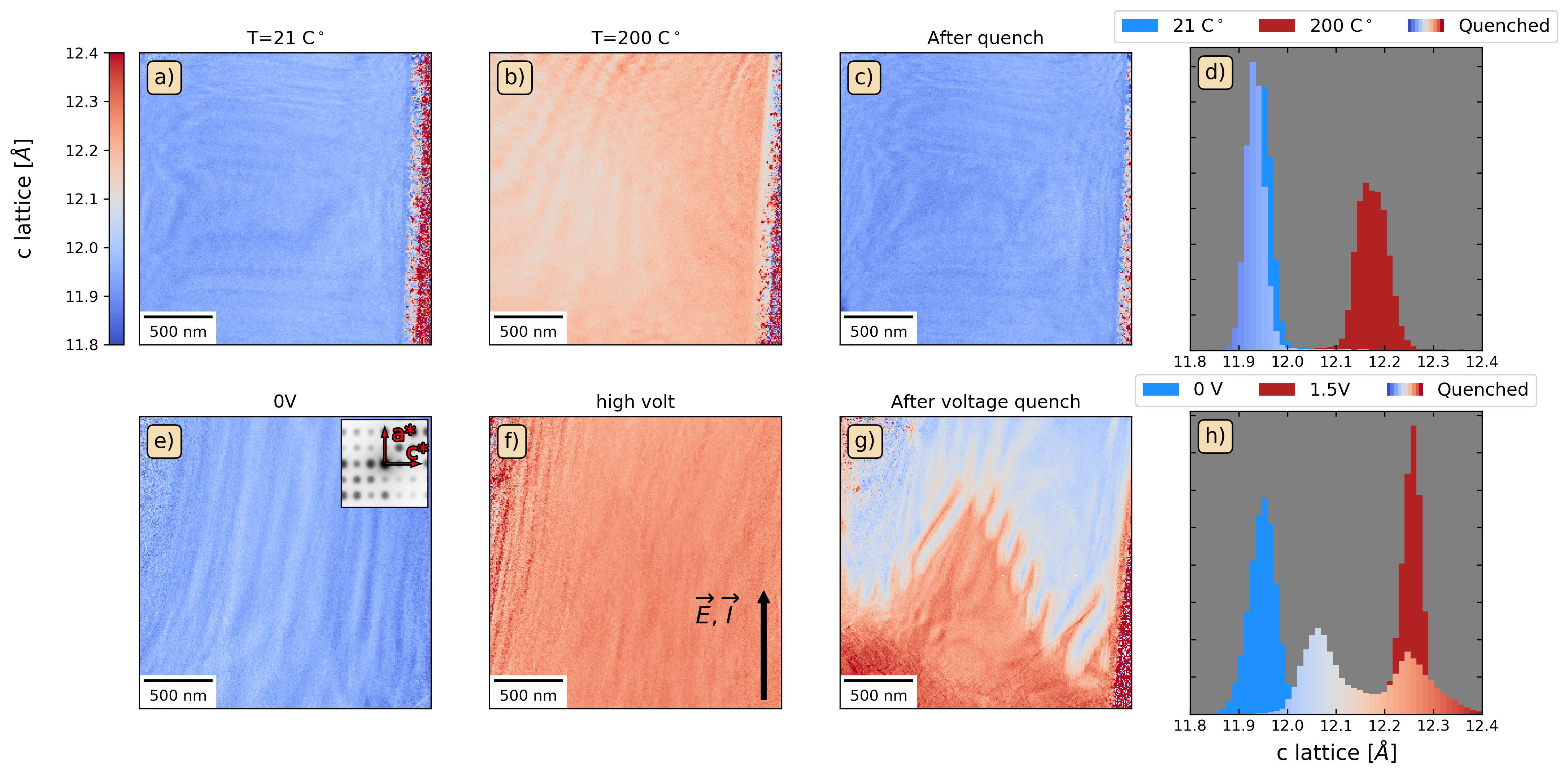}
\caption{\textbf{(a-c)} The real space map of the $c$ lattice parameter at room temperature, at 200°C and after quenching the specimen within 0.5~s back to room temperature. \textbf{(d)} Additionally, the histogram of the lattice parameter showing the two room temperature measurements and the high temperature measurement of the c-crystallographic axis. We can notice that with temperature alone the insulator-to-metal transition is fully reversible \textbf{(e-g)} The real space map of the $c$ lattice parameter at 0~V, at 1.5~V and after quenching back to 0~V of a specimen contacted to both electrodes with the field applied along the a axis (Joule heating is playing a role in this geometry as current can flow through the specimen). \textbf{(h)} The histogram which corresponds to the image in panel \textbf{(g)} where the colors of the bars correspond to the color in the image is displayed next to the right-most panel. Additionally, a histogram of the lattice parameter at 0 voltage \textbf{(e)} and maximum voltage \textbf{(f)} are shown to indicate the lattice constant in both insulating and metallic initial states. We notice that stripes are observed at the interface between two domains, one metallic and one with smaller lattice parameter (almost insulating) {{(see Supplmental Info for the I-V behavior before and after the voltage quench)}}. }
\label{fig:fig2b}
\end{figure*}

All those stripe patterns can be erased by switching on the voltage and bringing it to the highest probed amplitude necessary to have the uniform high voltage state, or heating the sample above the metal-insulator transition temperature T$_{\text{MI}}$. In both ways, the stripes can be systematically generated again with a similar spatial distribution when switching back the system to the zero voltage state. Hence, this patterned state is a stable configuration for the system which can be reproduced in a controlled manner. It is important to point out that, when considering a specimen attached to both sides of the capacitor, thus allowing a current flow through the sample, the crystallographic dependence of the pattern, as shown in Figure \ref{fig:fig2}, is suppressed. Indeed, by quenching the specimen from 200°C to room temperature (within 1 second) does not lead to the formation of any pattern (see Figure \ref{fig:fig2b} top panels). Similarly, by quenching the voltage to zero after bringing the specimen to the more conducting high-voltage configuration, no stripes are observed (see Figure \ref{fig:fig2b} bottom panels).  When electrical current is allowed to flow through the specimen, only fringes between a more conducting region (with elongated phase) and an insulating domain, similar to the results reported previously~\cite{Zhang2019}, are observed. We attribute this difference to Joule heating ~\cite{Mattoni2020}, which yields gradients of temperature in time during the temperature quench experiment that destabilize the orbital ordering. Hence, we argue that Joule heating or temperature gradients having a different relaxation time during the quench prevent the patterns formation. 

\begin{figure}[h!]
\centering
\includegraphics[width=0.35\textwidth]{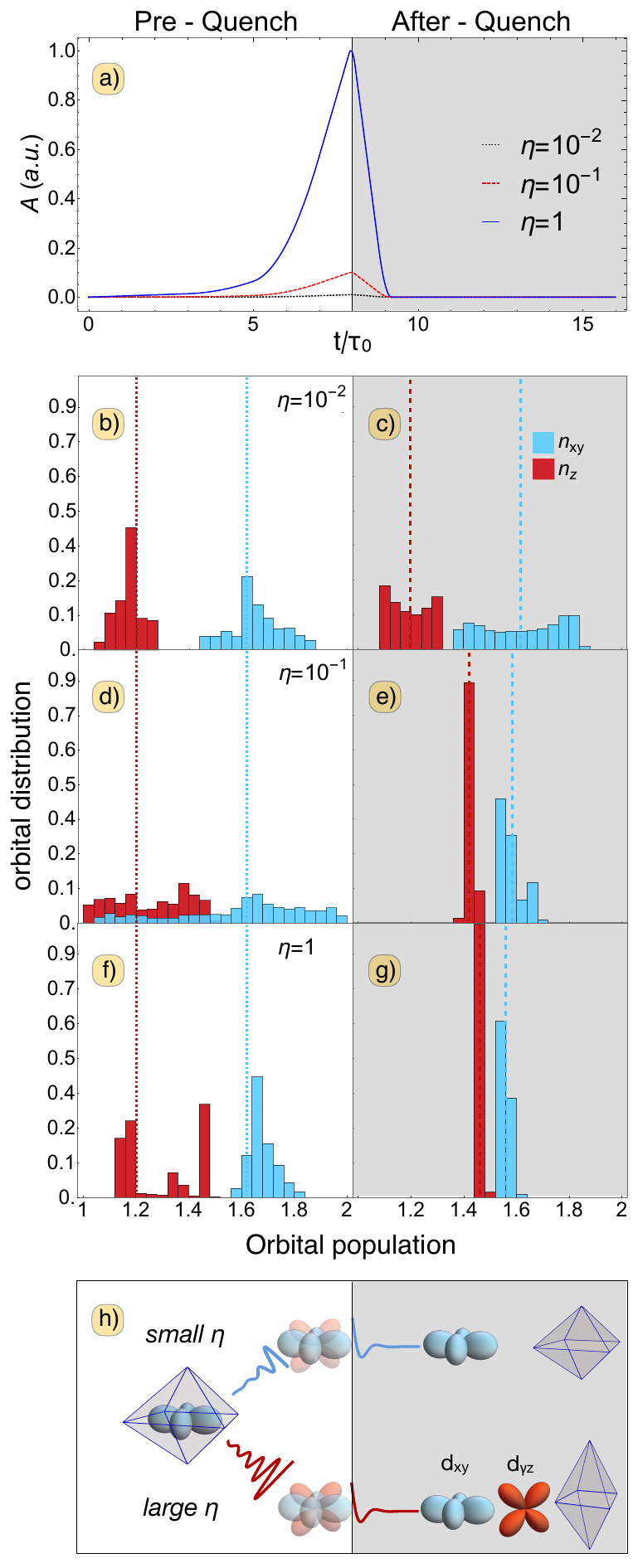}
\caption{\textbf{(a)} The electric field is introduced via a time dependent potential $A(t)$ as given by the Maxwell relation $E(t)=-\partial_{t}{A(t)}$ that is switched off after a characteristic time interval $\tau_0$. The $\eta$ parameter sets the strength of the electric field and it is defined as $E_{max}/E_{M}$, where $E_{max}$ is the maximum absolute value of $E(t)=-\partial_{t}{A(t)}$ and $E_{M} = 0.01 eV/{\rm{Angstrom}}$. Time is scaled in units of $\tau_0=100$ picoseconds. The details of the model parameters are reported in the Methods section.
\textbf{(b-g)} Time distribution of the density of the $d_{xy}$ and $d_{\gamma z}$ orbitals at the ruthenium site. The distribution is evaluated on the time interval preceding (left panels (b),(d),(f)) or following (right panels (c),(e),(g)) the quench of the electric field, for increasing values of electric field amplitude through $\eta$.
Vertical dotted lines mark the value of the orbital densities at zero voltage (left panels (b),(d),(f)), while dashed lines mark the values of the time average of the orbital densities after the quench (right panels (c),(e),(g)).
\textbf{(h)} Schematics of the evolution of the orbital population, before and after the quench, in the limit of weak (small $\eta$)  or strong electric field (large $\eta$).}
\label{fig:fig3}
\end{figure}

Let us focus on the nature and formation mechanism of the observed pattern.
Our first aim is to demonstrate that an orbital reconstruction can occur after the quench of the electric field. For this purpose, we perform a time-dependent simulation of the many body state on a finite size cluster with two effective RuO$_6$ octahedra. The analysis is aimed to capture on equal foot the correlated dynamics activated by the electric field both in space, on short-range length scale, as well as in time. The employed model Hamiltonian (see Supplemental Info) effectively includes all the local interactions at the Ru site and Ru-O charge transfer processes which are relevant to describing the correlated ground state in the CRO Mott phase. The electric field is introduced through a time-dependent vector potential that enters as a Peierls factor in the phase of the $d-p$ hopping amplitude (see Supplemental Information). For a quench dynamics it can be generally expressed by the profile in Figure \ref{fig:fig3}a.  

We start by considering the insulating state in the regime of flatten octahedra with the crystal field potential favoring the charge occupation of the planar $xy$ orbital. In this configuration, it is known \cite{Cuoco2006a,Cuoco2006b,Porter2018,DasPRX2018,Gorelov10} that there is an orbital unbalance with excess charge in the $xy$ band compared to the $(xz,yz)$ states, with direct correspondence with the octahedral distortions. We track the orbital dynamics in the time frame before and after the quench of the electric field for different amplitudes of the maximum applied electric field, $E_{max}$, expressed by the parameter $\eta=E_{max}/E_{M}$, with $E_{M} = 0.01 eV$ {{/$ \rm{Angstrom}$}}. 
{{$E_M$ is a reference scale of the amplitude of the electric field that, for convenience, we introduced when scanning the phase diagram in our simulation.
We have chosen the value 10$^{-2}$ eV/$\rm{Angstrom}$ because for the considered cluster, it is a characteristic scale that separates different regimes in terms of realized electronic configurations after the quench of the electric field.}}
For values of $\eta$ smaller than $10^{-2}$ there is no significant orbital variations in the time dynamics. Namely, the orbital occupation stays substantially unchanged with small fluctuations around the ground state values. For an applied electric field corresponding to $\eta\sim 10^{-2}$ the orbital dynamics starts to manifest significant fluctuations with a distribution having a broad variance around the ground state averaged occupation. For this regime of weak electric field the orbital unbalance is dynamically softened, although the averaged values are not much affected. The dynamics indicates that the charge and orbital distributions become broader in amplitude during the ramp up (Figure \ref{fig:fig3}b) and evolve into a different orbital distribution with larger spread (Figure \ref{fig:fig3}c) after the quench of the electric field.  
The increase of the maximal applied electric field before its switching-off has a significant impact on the orbital dynamics pre- and post-quench. The emergent orbital configuration is such that the orbital unbalance is completely suppressed (Figure \ref{fig:fig3}d) before the quench, while after the quench (Figure \ref{fig:fig3}e) the distribution shrinks in amplitude and the difference of the averaged charge occupation in the $xy$ and $(xz,yz)$ states tends to vanish. A further increase of the maximal electric field amplitude (i.e. $\eta\sim 1$) is affecting the profile of the orbital distribution before the quench while keeping the qualitative trend of suppressing the charge separation in the $xy$ and $(xz,yz)$ orbital sectors. 

Hence, the analysis demonstrates that the application of an electric field yields non-equilibrium orbital configurations that are compatible with short and long octahedra depending on the parameter $\eta$. After the quench, above a critical threshold of the electric field, the orbital unbalance is substantially suppressed thus favoring the formation of more elongated octahedra. This result implies that, due to the presence of spatial electric field gradients, the system tends to rearrange by forming interfaces between long and short octahedra.

\begin{figure}[h!]
\centering
\includegraphics[width=8.5cm, angle=0]{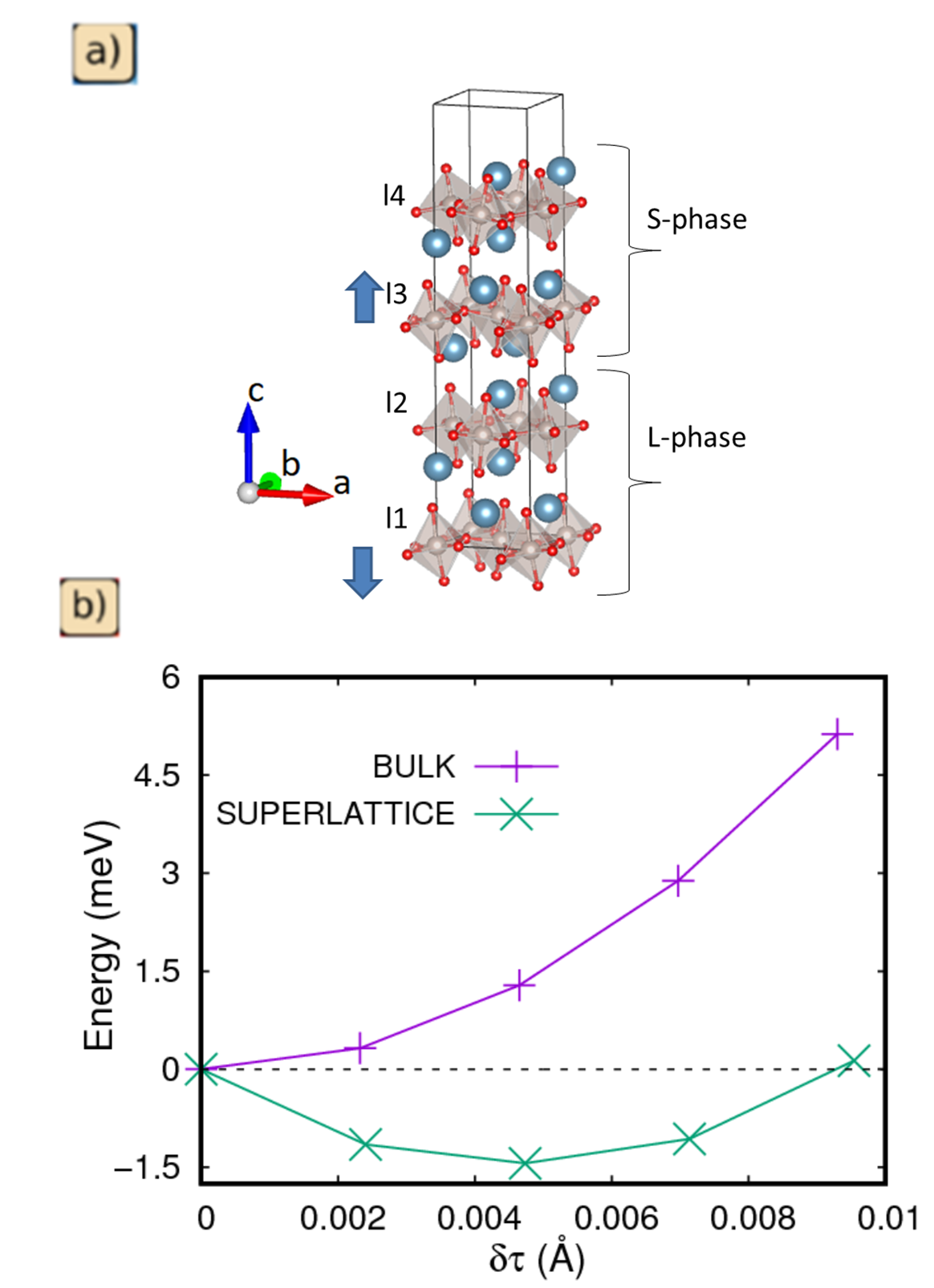}
\caption{\textbf{(a)} Superlattice of CRO composed of four RuO$_2$ layers. Two layers are in the L- and two in the S-phase. Grey, red and blue spheres indicate the Ru, O and Ca atoms, respectively. The blue arrows indicate the displacements, $\delta\tau$, of Ru atoms as due to structural relaxation. $l1$, $l2$, $l3$ and $l4$ label the layers in the superlattice in the $L$- and $S$-regions.
\textbf{(b)} Energies of the superlattice and the bulk as a function of displacements $\delta\tau$ of the Ru atoms with respect to the centrosymmetric positions.}
\label{fig:fig4}
\label{Superlattice}
\end{figure}

Having established that the quench of the electric field results into a {{reduction of the orbital unbalance that is compatible with long unit cell}}, we consider how the formation of interfaces between long and short octahedra interacts with the electric field. For this purpose, we simulate a superlattice configuration with long (L) and short (S) CRO unit cells (Fig. \ref{fig:fig4}a). We determine the optimized structure by the relaxation of the Ru atoms along the c-axis and we obtain a head-to-head displacement of the Ru layers, as reported in Fig. \ref{Superlattice}a. This configuration implies that the overall displacement is vanishing. The Ru layers at the interface between the L- and the S-phase move towards the S-phase. Notably, qualitative similar results are observed for supercells of different size.
We calculate the free energies of the superlattice and the bulk as a function of the displacements of the Ru atoms with respect to their centrosymmetric positions.

As we can see from Fig. \ref{fig:fig4}b, the equilibrium energy of the bulk is achieved when the Ru atoms are in the centrosymmetric positions, while for the superlattice the situation changes, namely the equilibrium energy is when the Ru atoms are displaced by approximately 0.47~pm with respect to their centrosymmetric positions. These shifts of the Ru atoms along the (001) direction at the interface between different Ru-based compounds stacked along (001) are similar to those predicted in the metallic phase of the Sr-based ruthenate compounds \cite{Autieri2014}. While larger displacements are found in the metallic phase, we expect that the size of the displacements mainly depends on the electronic mismatch between the two structurally distinct phases within the superlattice. These displacements make the interface electrically active {{because they can sustain a non-vanishing electric dipole}}. Hence, the resulting physical scenario is that the application of an electric field activates the orbital and lattice dynamics which allows for deviations of the structural configurations from the equilibrium. When interface configurations with inequivalent octahedral distortions form, the system tends to stabilize them and lower the energy by having a periodic alternation of short and long unit cells (Fig. \ref{fig:fig4}). The head-to-head interface configuration is compatible with an averaged net vanishing electric field. This behavior resembles the phenomenology of magnetic systems described by the kinetic Ising model \cite{Bray1994} with the orbital-lattice degrees of freedom replacing the spin ones.

To wrap up, we have demonstrated that the ramp up to a critical amplitude of the applied voltage and its successive quench to zero are able to induce a stripe phase in the CRO Mott insulator. The stripe phase is marked by long and short octahedra that periodically alternate along the $c$-axis {{with a nanometric length scale}}. 
{{This pattern together with the way (i.e. gate voltage quench) it is generated mark the difference between the observed stripe phase and the stripe domains realized at the boundary of micrometric regions with metallic and insulating character \cite{Vitalone2022}}}.
The configuration is stable and can be controlled by varying and switching off the applied gate. The stripe formation mechanism depends on the orientation of the applied electric field thus underlining the role of the orbital degrees of freedom in the stripe phase formation. 
{{We argue that the reason for having a different response for an electric field oriented along the $a$ and $b$ axes might arise from the character of the activated orbital excitation due to presence of an orbital easy axis for the orthorhombic crystalline symmetry of CRO.}}

Besides, in the insulating phase the ground state is orbitally and structurally correlated. The electric field tends to destroy the orbital pattern by deforming the octahedra and brings the system into a new state with interfaces between orbitally and structurally inequivalent configurations. The interfaces carry an electric dipole, so that they can interact and stabilize a periodic arrangement in the form of stripes. 
The observed phenomena may have a high impact on innovative types of switching memories, once the stripes pattern has been written, and they can be erased and written at will just by applying a small amplitude voltage. The fact that the stripe phase becomes the new zero-voltage state is also very beneficial as it is more energy efficient than having to switch back to a fully insulating state. These findings thus can pave the way for the construction of low-energy consumption non-volatile nanoscale electronics and in perspective be integrated with other functional devices employing photonic effects.


{\textbf{Acknowledgement}}
This project has received funding from the European Union's Horizon 2020 research and innovation programme under grant agreement No 823717 – ESTEEM3. The Merlin camera used in the experiment received funding from the FWO-Hercules fund G0H4316N 'Direct electron detector for soft matter TEM'.
C. A. and G. C. are supported by the Foundation for Polish Science through the International Research Agendas program co-financed by the
European Union within the Smart Growth Operational Programme.
C. A. and G. C. acknowledge the access to the computing facilities of the Interdisciplinary
Center of Modeling at the University of Warsaw, Grant No.~GB84-0, GB84-1 and GB84-7 and GB84-7 and Poznan Supercomputing and Networking Center Grant No. 609..
C. A. and G. C. acknowledge the CINECA award under the ISCRA initiative IsC85 "TOPMOST"
Grant, for the availability of high-performance computing resources and support. We acknoweldge A. Guarino and C. Elia for providing support about the electrical characterization of the sample. M.C., R.F., and A.V. acknowledge support from the EU’s Horizon 2020213
research and innovation program under Grant Agreement No. 964398 (SUPERGATE).



\section{Supplemental Info}

In the Supplemental Information we provide details about the experimental methods and setup, the real space maps for various electric field configurations, and the electrical-structural characterization. Furthermore, we describe the methodology for the simulation of the electric field quench, and aspects of the density functional theory employed to investigate the superlattice.

\section{Experimental methods}


In this Section we describe the methods related to the crystal fabrication, the structural characterization, and the preparation and the analysis for transmission electron microscopy.

\subsubsection{Fabrication and structural characterization}
Single crystals were grown by the floating zone technique using an infrared image furnace with two mirrors (NEC Machinery, model SC1-MDH11020). CRO single crystals used in this experiment were carefully selected prior to HRSTEM analysis. The morphology and composition were inspected by scanning electron microscopy using a SEM Leo EVO 50 Zeiss, coupled with an energy dispersive spectrometer (Oxford INCA Energy 300). The structural characterization was performed by high-angle X-ray diffraction measurements using a Panalytical X-Pert MRD PRO diffractometer and the electrical characterization was obtained by two terminal method applying DC current along the c-axis of the crystal at room temperature (see supplementary material)

\subsubsection{Specimen preparation for Transmission Electron microscopy}
Cross-sectional cuts of the samples along the [100] and [010] directions of Ca$_2$RuO$_4$ c-oriented single crystal were prepared using a Thermofisher Scientific Helios 650 dual-beam Focused Ion Beam device on dedicated DENS biasing chips as shown in the supplementary material. To get a sample with field applied along the c-crystallographic axis, the lift-out lamella was rotated by 180° using the omniprobe nanomanipulator before attaching it to the chip this resulted in a slight angle between the electrical field and the c-axis of the crystal which has been neglected. {{The sample thickness was kept around 100-150 nm thick}}. Biasing and heating experiments were carried out in a DENS Solutions Lightning double tilt holder with help from a Keithley 2400 Source Meter and in-house control program. 

\subsubsection{Scanning Transmission Electron microscopy}
The electron microscopy characterization was performed on the X-Ant-Em instrument at the University of Antwerp. The Electron Microscope used consists of an FEI Titan G3 electron microscope equipped with an aberration corrector for the probe-forming lens as well as a high-brightness gun operated at 300~kV acceleration voltage with a beam current of around 100~pA for all experiments to reduce acquisition time. The STEM convergence semi-angle used was 21~mrad for HRSTEM-HAADF imaging, providing a probe size of ~0.8 Å. The collection semi-angle ranges from 29-160~mrad for annular dark field (ADF) imaging.\\
Diffraction patterns used for Fig. 1 (main text) were acquired in nano-beam electron diffraction(NBED) mode with a convergence angle of 0.25~mrad, resulting in a spatial resolution of $\sim$ 1~nm, and a collection angle of 21~mrad using a camera length of 285~mm and a 256$\times$256 pixel Quantum Detectors Merlin direct electron detection camera with an acquisition time of 2ms/pixel. Similar conditions were used to acquire the 2D maps presented in Fig. 2 and Fig. 2b (main text).

\subsubsection{Determination of the lattice parameters using HRSTEM-HAADF (direct space).}
The HR-STEM images were used to determine the lattice parameters of the CRO crystal. Ten frames were acquired with a dwell time of 2~$\mu$s. Since each individual image contains enough signal it is possible to align them with the cross correlation method \cite{Savitzky2018}. Multiple fast scans were acquired to reduce the effect of the sample drift while retaining the same signal-to-noise as doing one slow scan. After the images were aligned, a peak finding routine implemented in Statstem \cite{DeBacker2016} was used to extract the initial guess of the atomic positions. These initial positions were refined by fitting 2D Gaussians to each atomic column. The fitted values of the centre were used to determine the lattice constants of the material.

\subsubsection{Determination of the lattice parameters using NBED (reciprocal space)}
For NBED, a diffraction pattern is acquired at each probe position making it possible to map the lattice parameter at each probe position. In order to do this, a local 2D peak finding algorithm is used to determine the position of each diffraction peak  \cite{Nord2020}. Once these positions are determined, the two lattice vectors which describe the diffraction peak positions is determined by performing a linear fitting procedure. Once the lattice vectors are determined it is straightforward to retrieve the norm of each vector which corresponds the length of the lattice parameters.

\newpage

\section{Experimental setup and electrical-structural characterization}

In this Section we present extra results about the experimental setup (Fig. S1), the real space maps for various electric field configurations (Fig. S2), and the electrical-structural characterization (Fig. S3).

\renewcommand{\figurename}{{Figure S}}

\begin{figure*}
\includegraphics[width=1.\textwidth]{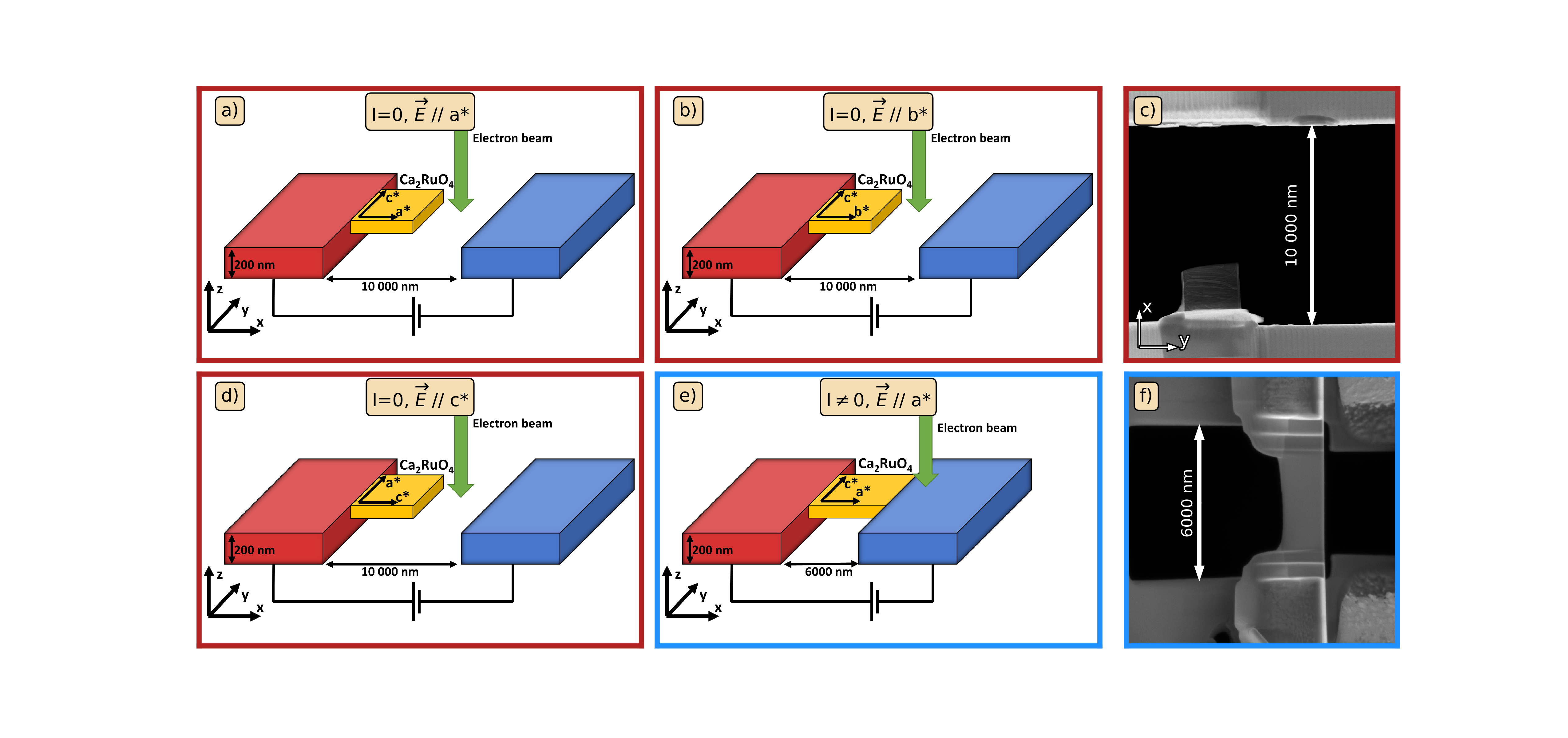}
\caption{\textbf{(a,b,d)} A sketch of the experimental setup where the sample is connected to one electrode. A voltage is then applied over the two electrode which creates an electric field along the a, b and c crystallographic axes respectively. \textbf{(c)} An overview scan of the pure electric field setup where the sample is clearly visible attached to only one electrode (geometry used in Fig. 1 (g-h), Fig. 2 and Fig. 4 \textbf{(e)} (main text). A sketch of the experimental setup where the sample is connected to one electrode allowing no current to flow through the specimen. \textbf{(f)} An overview scan of the current allowed setup where the sample is clearly visible attached to both electrodes (related to {Extended Figure 2} (e-h)).}
\label{fig:figS1}
\end{figure*}

\begin{figure*}
\includegraphics[width=1.\textwidth]{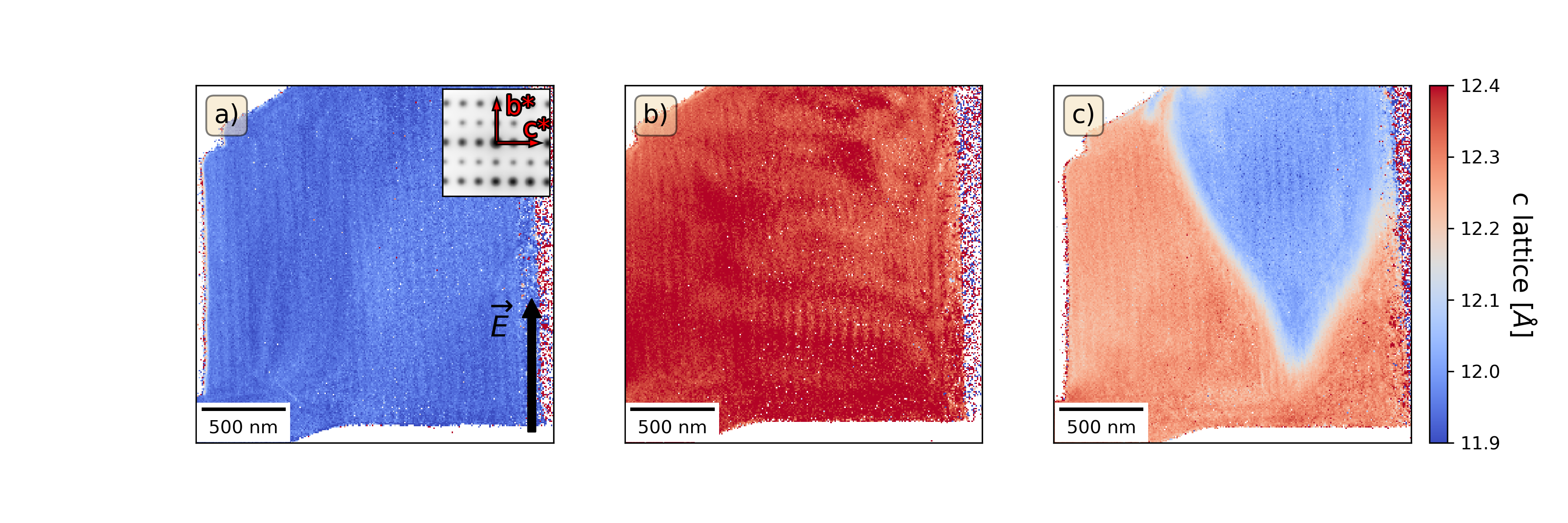}
\caption{The real space maps of the $c$ lattice parameter when applying field along the b axis[corresponding to
Fig. 2(a) (main text) and Fig. S1(b)]\textbf{(a)} before applying an electric field of the sample with E-field along the b axis \textbf{(b)} The
real space map of the $c$ lattice parameter while applying the maximum electric field \textbf{(c)} The real space map of the $c$ lattice
parameter after quenching the specimen back to 0V.}
\label{fig:figS2}
\end{figure*}

\begin{figure*}
\includegraphics[width=\textwidth]{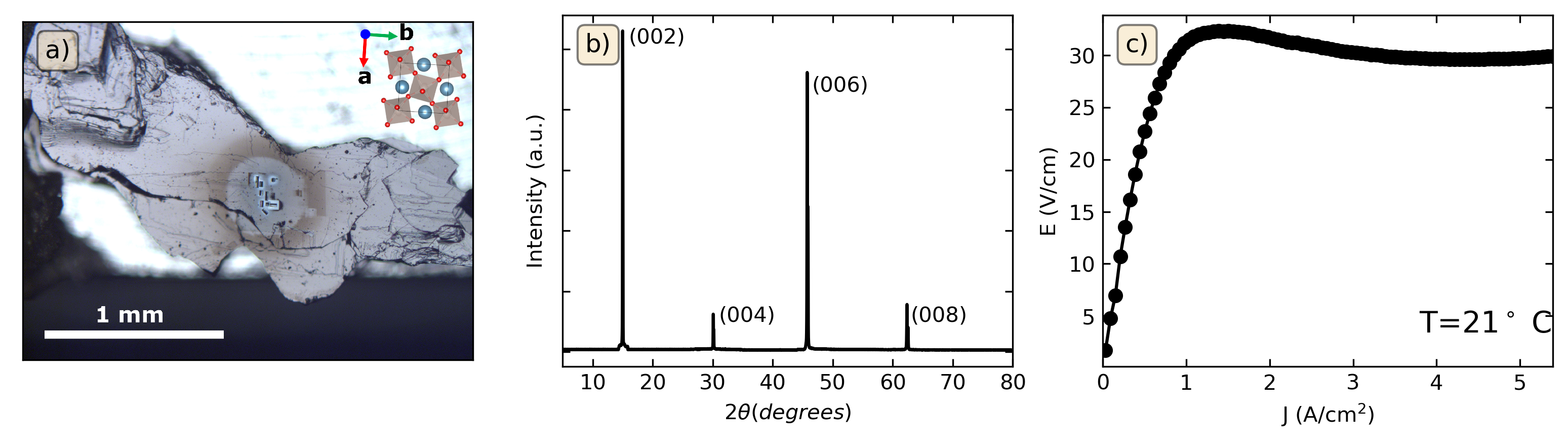}
\caption{\textbf{(a)} Top view optical image of the specimen used for the STEM measurements. In the center we can notice the locations where the FIB-lamella were extracted along both a- and b-crystallographic axis, the inset shows the in-plane orientation of the crystal determined through EBSD measurement.\textbf{(b)} High-angle X-ray diffraction pattern acquired on the single crystal shown in panel (a) showing the absence of impurity peaks.\textbf{(c)} Electric characterization of the single crystal shown in panel (a) representing the measured electric field (E in V/cm) against the applied current density (J in A/cm$^2$) as taken at room temperature}
\label{fig:Graph1}
\end{figure*}

\begin{figure*}
\centering
\includegraphics[width=8.5cm, angle=0]{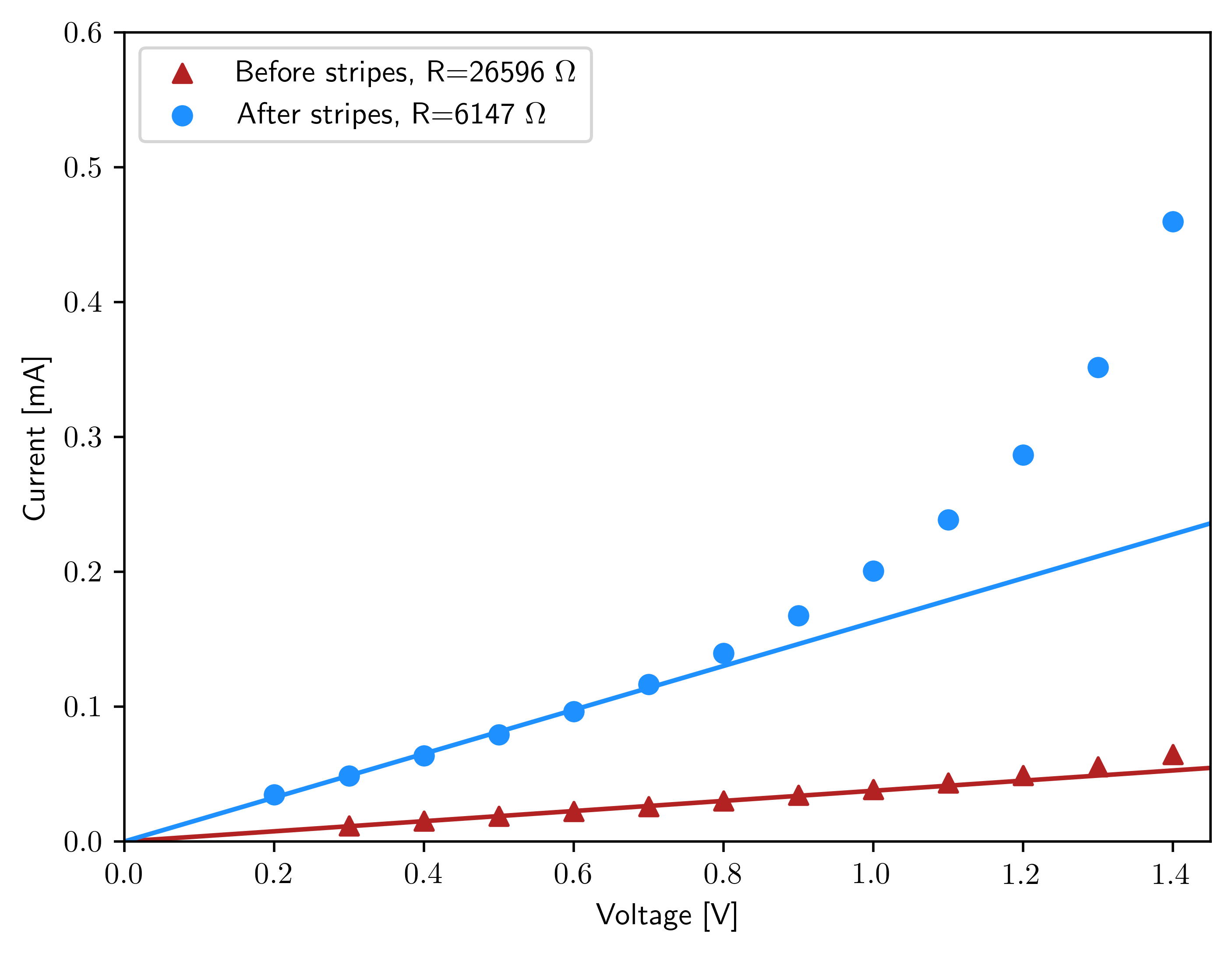}
\caption{Current-voltage (I-V) curves of the sample with a contact configuration as in Fig. S1e. The measurement is performed before (triangle) and after (circle) establishing the stripe phase corresponding to the pattern in Figure 3g of the main text. The amplitude of the resistance extracted from the linear regime is also reported in the label.}
\label{Figure R1}
\label{}
\end{figure*}

\newpage

\section{Theoretical modelling of the electric field quench}

In CRO, the bands close to the Fermi level stem mostly from the $t_{2g}$ orbitals $(d_{yz,},d_{zx},d_{xy})$, which hybridize with the oxygen $(p_x,p_y,p_z)$ bands. Hence, one can build an effective model Hamiltonian for the propagating electrons within the ruthenium-oxygen plane, by considering the interaction terms at the ruthenium and oxygen sites and the kinetic term responsible for the ruthenium-oxygen hybridization.\\
The non-interacting part of the Hamiltonian for the Ru-O bond along the $x$ ([100]) direction comprises the following terms:
\begin{eqnarray}
H_{Ru_i-O}[x] &=& t_{d_{\alpha},p_{\beta}}\left[ d_{i,\alpha\sigma }^{\dagger} p_{\beta\sigma
}+h.c.\right] \label{eq:Hkin} \\
H_{el}^O &=& \varepsilon _{x} n_{p_x}+\varepsilon _{y}
n_{p_y}+\varepsilon_{z} n_{p_z} \label{eq:HelO}\\
H_{el}^{Ru} &=&\sum_{i} \varepsilon_{yz}n_{id_{yz}}+\varepsilon_{zx}n_{id_{zx}}+\varepsilon_{xy}n_{id_{xy}}\ .
\label{eq:HelRu}
\end{eqnarray}

\noindent  Eq.~(\ref{eq:Hkin}) is the Ru-O hopping along a given symmetry direction, e.g. the $x$-axis, $t_{d_{\alpha},p_{\beta}}$ is the hopping amplitude, $\alpha,\beta $ are orbital  indices running over the three orbitals in the t$_{2g}$
sector, and $d_{i\alpha \sigma }^{\dagger }$ is
the creation operator of an electron with spin $\sigma $ at the site $i$ in
the orbital $\alpha$. Here, we include all the hopping terms which are allowed according to the Slater-Koster rules, assuming that the Ru-O bond can form an angle $\theta$ with the $x$ axis, due to the rotation of the octahedra around the $c$-axis.
Eqs.~(\ref{eq:HelO}) and (\ref{eq:HelRu}) describe the orbital dependent on-site energy terms, which take into account the offset between the occupied orbitals of O and Ru.  In particular, Eqs~(\ref{eq:HelRu})  includes the on-site crystal-field splitting of the $t_{2g}$ manifold in the octahedral environment, which can be expressed in terms of the amplitude $\Delta_{CF}$, with $\Delta_{CF}=(\varepsilon_{xy}-\varepsilon_z)$. For flat octahedra below the structural transition temperature of Ca$_"$RuO$_4$, $\Delta_{CF}$ is negative. We also consider the possibility of having a small orthorhombic splitting, $\delta_{ort}$ of the $d_{xz}$,$d_{yz}$ orbitals by assuming that $\varepsilon_{yz}=\varepsilon_{z}+\delta_{ort}$ and $\varepsilon_{yz}=\varepsilon_{z}-\delta_{ort}$.\\
For interacting electrons, we limit to the local Hamiltonian $H_{\text{el-el}}^{Ru}$,\cite{Cuoco2006a,Cuoco2006b,Pincini2019} at the Ru sites, which includes the complete Coulomb interaction projected onto the t$_{2g}$ subspace. This is given by the intra-orbital $U$, and the inter-orbital Coulomb and
exchange elements, $U^{^{\prime }}$ and  $J_{\mathrm{H}}$. We assume a rotational invariant condition for the Coulomb amplitudes, so that $U=U^{^{\prime }}+2J_{H}$, and $J^{^{\prime}}=J_{H}$
\begin{eqnarray}
H_{\text{el-el}}^{Ru} = &&
U\sum_{i,\alpha} n_{i\alpha \uparrow }n_{i\alpha \downarrow}
-2J_{\mathrm{H}}\sum\limits_{i,\alpha <\beta }\bf{S}_{i\alpha }\cdot 
\bf{S}_{i\beta } +\\ &&
+\left(U^{^{\prime }}-\frac{J_{\mathrm{H}}}{2}\right)\sum\limits_{i,\alpha <\beta }n_{i\alpha }n_{i\beta } +\nonumber \\ &&+J_{H}\sum\limits_{i,\alpha <\beta }d_{i\alpha \uparrow }^{\dagger }d_{i\alpha
\downarrow }^{\dagger }d_{i\beta \uparrow }d_{i\beta \downarrow }.
\end{eqnarray}

\noindent Moreover, we consider the spin-orbit coupling $H_{\mathrm{SOC}}^{Ru}$

\begin{eqnarray}
H_{\mathrm{SOC}}^{Ru}&=&\lambda \sum\limits_{ i \alpha ,\sigma }\sum_{\beta ,\sigma
^{^{\prime }}} d_{i\alpha \sigma }^{\dagger } ({\bf l}_{\alpha\beta } \cdot {\bf S}_{\sigma \sigma ^{^{\prime }}})
d_{i \beta
\sigma ^{^{\prime }}}\,
\end{eqnarray}
where $\lambda$ is the spin-orbit coupling strength and $ (\bf{l}_{\alpha\beta }\cdot {\bf{s}_{\sigma \sigma ^{^{\prime }}}})$ are the matrix elements of the atomic SOC in the $t_{2g}$ basis. Note that the $t_{2g}$ orbitals have an
effective orbital momentum $l = 1$, whose components in the basis $(d_{yz},d_{xz},d_{xy})$ can be expressed as $l_k=i \varepsilon_{kpq}$.
\\
For the examined cluster with two rutheniums ions  Ru$_1$ and Ru$_2$ and one oxygen atom O, the total Hamiltonian definitely reads as:
\begin{equation}
H=H_{Ru_1-O}[x]+H_{Ru_2-O}[x]+H_{el}^O+H_{el}^{Ru}+H_{el-el}^{Ru}+H_{\mathrm{SOC}}^{Ru}\ .
\end{equation}
For the present analysis we adopt material specific
values such as $\lambda=0.075$~eV, $U$ in the range [2.0,2.2]~eV, and $J_{H}$ [0.35, 0.5]~eV are taken as a reference for the analysis. Similar values for $\Delta_{CF}$, $U$ and $J_H$ have been used for calculations of electronic spectra in CRO and the ratio $g=\Delta_{CF}/(2 \lambda)$ is typically considered to lie in the range $\sim$[1.5,2] for modelling the spin excitations observed by neutron scattering\cite{Sutter2017,Veenstra2014,DasPRX2018,Jain2017}. For the hopping amplitudes, we assume that the basic $p-d$ hopping amplitudes in the tetragonal ($\alpha=0$) symmetry have the following value $t^{0}_{p,d}$=1.5~eV.\\

Let us describer the methodology for investigating the consequence of a time-dependent electric field that is switched off after a given time interval.
In the presence of an applied voltage, the effect of the external electric field can be incorporated in the miscroscopic model by the standard Peierls substitution to the hopping matrix
elements,
\begin{eqnarray}
t_{d_{\alpha},p_{\beta}}(t)=t_{d_{\alpha},p_{\beta}} \exp\left[-i\,\frac{e}{\hbar} \int_{r_{Ru}}^{r_{O}} {\bf{A(t)}} d{\bf r} \right]
\end{eqnarray} 
\noindent where $r_{O}$ and $r_{Ru}$ are the position of the O and Ru atoms, $e$ is the electron charge and $\hbar$ the Planck constant, while the vector potential is related to the electric field by $\bf{E}(t)=-\partial_{t}{\bf A(t)}$. The electric field in this formalism corresponds to a time-dependent deformation of the Hamiltonian and the present approach avoids to deal with electrodes in the system. Assuming that the electric field is static lying in the Ru-O plane and taking only one projection along the Ru-O-Ru axis, one can describe the evolution of the ground state by introducing a scalar vector potential $A(t)$. We model the quench behavior of the electric field, by assuming the time profile for $A(t)$ displayed in Fig. 4(a) (main text). In the time interval preceding the quench  $t<t_Q$, $A(t)$ grows from zero to a maximum value, showing a super-linear dependence in time. In the specific, it is obtained as a cubic polynomial interpolation between linear functions, where the strength of the electric field in gradually increased up to an absolute value $E_{max}$.  In order to explore different coupling regimes compatible with the experimental values of the applied voltage, we considered several cases which are parametrized by the constant $\eta=E_{max}/E_{M} $, with $E_{M} = 0.01 eV/\text{\AA}$ and $E_{max} $ in the range [$10^{-4}$,$10^{-2}$]  $\text{eV}/\text{\AA}$.
At $t_Q$ of the order of 0.8 ns, $A$ is suddenly reduced to zero over a time interval of 0.1 ns.\\
From a methodological point of view, we need to solve the time-dependent Schr{\"o}dinger equation, $i \hbar \frac{\partial}{\partial t} |\Psi(t)\rangle = H(t) |\Psi(t)\rangle$, which rules the time evolution of the quantum system at zero temperature starting from the ground state of the Hamiltonian obtained by exact diagonalization. Due to the large dimension of the Hilbert space, the time dynamics of the many-body ground state is performed by means of the Cranck-Nicholson's method, that guarantees a unitary time evolution where the evolved wave function is expressed in an infinitesimal interval as 
\begin{widetext}
\begin{eqnarray}
|\Psi(t+\Delta t)\rangle=\exp[-i \hbar^{-1} \int_{t}^{t+\Delta t } H(t) dt] |\Psi(t)\rangle \nonumber \approx \frac{[1-i \frac{\Delta t/2}{\hbar} H(t+ \Delta t/2)]}{1+i \frac{\Delta t/2}{\hbar} H(t+ \Delta t/2)] }  |\Psi(t)\rangle \,.
\label{eq:CrankNich}
\end{eqnarray}
\end{widetext}
Hence, by means of Eq. (\ref{eq:CrankNich}) we determine the time dependent evolution of the ground state by discretizing the time interval. Here, the time step is considered to be $dt=1.0\times 10^{-2} \hbar/t^{0}_{p,d}$, with $t^{0}_{p,d}$ being the amplitude of the p-d $\pi$ hybridization hopping process for $\theta=0$. The choice of the time step is small enough to guarantee the convergence for the solution.\\
Finally, we provide the description of the out-of-equilibrium dynamics of the on-site orbital occupancy of the $d$-orbitals in the ground-state following the quench, by calculating the time dependent expectation value of the electron density $n_{xy}$ in the $d_{xy}$, and averaged ($d_{xz}$,$d_{yz}$) orbitals as given by $\frac{1}{2}(n_{xz}+n_{yz})$, respectively. These quantities are the most relevant to identify the modification of the orbital configuration after quenching the applied electric field.

\section{Density functional theory for CRO superlattice}

We have performed DFT calculations by using the Vienna ab-initio simulation package (VASP) \cite{Kresse93,Kresse96a,Kresse96b}. The core and
the valence electrons were treated within the projector augmented
wave (PAW) \cite{Kresse99}  method with a cutoff of 480~eV
for the plane-wave basis. We have used the PBEsol
exchange-correlation method \cite{Perdew08}, a revised Perdew-Burke-Ernzerhof (PBE) that improves equilibrium properties of solids.
PBEsol+$U$ is the approach that we have followed to take into account the correlations associated with the Ru-4$d$ states.
We have considered $U$=3 eV for the antiferromagnetic insulating phase of ruthenates\cite{Autieri_2016,Gorelov10}, and regarding the Hund coupling we have used the value
$J_H$ = 0.15 $U$ in agreement with approaches based on the constrained
random phase approximation for $4d$-electrons \cite{Vaugier12}.
The values of the lattice constants are $a_S$=5.3945~{\AA}, $b_S$=5.5999~{\AA}, $c_S$=11.7653~{\AA} in the S-Pbca phase and $a_L$=5.3606~{\AA}, $b_L$=5.3507~{\AA}, $c_L$=12.2637~{\AA} in the L-Pbca phase \cite{Friedt01}. To simulate the stripe phase, we built a superlattice composed of four RuO$_2$ layers, with two layers in the L- and two layers in the S-phase stacked along the $c$-axis.
The lattice constants of the superlattice are obtained by averaging the lattice parameters of the bulk; we have that $a_{superlattice}$=($a_S$+$a_L$)/2, $b_{superlattice}$=($b_S$+$b_L$)/2 and $c_{superlattice}$=$c_S$+$c_L$.
A 11$\times$11$\times$4 k-point grid has been used for the bulk~\cite{Zhang2017}, while a 11$\times$11$\times$2 k-point grid has been used for the superlattice.\\

\providecommand{\latin}[1]{#1}
\makeatletter
\providecommand{\doi}
  {\begingroup\let\do\@makeother\dospecials
  \catcode`\{=1 \catcode`\}=2 \doi@aux}
\providecommand{\doi@aux}[1]{\endgroup\texttt{#1}}
\makeatother
\providecommand*\mcitethebibliography{\thebibliography}
\csname @ifundefined\endcsname{endmcitethebibliography}
  {\let\endmcitethebibliography\endthebibliography}{}

\end{document}